\documentclass{article}

\usepackage{PRIMEarxiv}

\usepackage[utf8]{inputenc} % allow utf-8 input
\usepackage[T1]{fontenc}    % use 8-bit T1 fonts
\usepackage{hyperref}       % hyperlinks
\usepackage{url}            % simple URL typesetting
\usepackage{booktabs}       % professional-quality tables
\usepackage{amsfonts}       % blackboard math symbols
\usepackage{nicefrac}       % compact symbols for 1/2, etc.
\usepackage{microtype}      % microtypography
\usepackage{lipsum}
\usepackage{fancyhdr}       % header
\usepackage{graphicx}       % graphics
\usepackage{subfigure}
\graphicspath{{media/}}     % organize your images and other figures under media/ folder

\title{Electrified Autonomous Freight Benefit analysis on Fleet, Infrastructure and Grid Leveraging Grid-Electrified Mobility (GEM) Model}

\author{
Wanshi Hong\\
Lawrence Berkeley National Laboratory\\
Berkeley, CA, USA\\
\texttt{wanshihong@lbl.gov} \\
\And
Alan Jenn\\
University of California, Davis\\
Davis, CA, USA\\
\texttt{ajenn@ucdavis.edu} \\
\And
Bin Wang\\
Lawrence Berkeley National Laboratory\\
Berkeley, CA, USA\\
\texttt{wangbin@lbl.gov} \\
}

\begin{document}
\maketitle

\begin{abstract}
This paper analyzes the potential benefit of heavy-duty vehicle (HDV) electrification and automation on fleet cost, infrastructure cost, grid, and environmental impact. In this work, we extended the vehicle electrification benefit analysis tool: Grid-Electrified Mobility (GEM) model, which had primarily been used to study light-duty passenger vehicles (LDVs), to analyze the heavy-duty vehicle electrification. The extended model is derived for freight transportation and key results and findings on the impact of freight electrication and autonomization are presented and discussed.
\end{abstract}

% keywords can be removed

\section{Introduction} \label{sec1}
The transportation sector is undergoing a transformation through the introduction of on-demand mobility and vehicle automation thanks to a variety of emerging mobility technologies \cite{greenblatt_automated_2015}. These advances, combined with electrification, could create new synergies that would provide high-quality, low-cost, and energy-efficient mobility at scale \cite{fulton_three_2018}. However, the adoption of plug-in electric vehicles has been relatively slow for several reasons, including technological uncertainty, slow charging, range anxiety, and higher capital costs than other types of vehicles \cite{green_increasing_2014}. This is especially true in the freight industry, particularly around heavy-duty truck electrification and operations. As major truck fleet operators and truck manufacturers have announced plans to accelerate truck electrification, filling these gaps in system modeling capabilities will be crucial. For example, Walmart aims to electrify its entire truck fleet within a decade \cite{stumpf_walmart_2018}. The uptake in the adoption of electric trucks is important in the context of rising freight demand, which is projected to grow by 52\% from 397 billion miles in 2018 to 601 billion miles in 2050 projected by U.S. EIA \cite{noauthor_u.s._nodate}. While there is still a great deal of uncertainty around the exact impact that automated vehicles will have on the transportation system in the coming decades \cite{stephens_estimated_2016}, many believe that they could soon become a substantial part of the transportation system, dramatically disrupting conventional modes of mobility. 

Overall, the urgent need to decarbonize the transportation sector combined with falling battery prices has spurred industry and policy interest in long-haul truck electrification. Understanding the charging behavior and resulting loads from freight electrification will be critical for the smooth operation of the electric grid and will have far-reaching impacts on the environment in the form of greenhouse gas (GHG) emissions and air pollution. As such, this work has aimed to assess the benefits of heavy-duty truck electrification and emerging vehicle electrification opportunities in micro-mobility markets using the Grid-Integrated Electric Mobility Model (GEM) and Medium and Heavy-Duty Electric Vehicle Infrastructure - Load Operations and Deployment (HEVI-LOAD). This national model simultaneously optimizes the provision and operation of shared heavy-duty autonomous electric vehicles (SHAEVs) to provide electrified goods mobility alongside an economic dispatch of power generation \cite{tong_energy_2021}.

Increasing levels of renewable energy are being added to the electric grid while vehicle electrification is on the rise. The impacts of integrating these technologies require new analytical methodologies that couple capabilities across the transportation and power sectors. This work has further developed the GEM model to explore the dynamics and impacts of an integrated intelligent transportation–grid system in which mobility is served by either privately owned electrified trucks or SHAEVs, charging is responsive to costs on the grid, and power resources are dispatched in merit order to serve electricity demand. 

\subsection{Objective}
The objective of this work includes:
\begin{itemize}
    \item Develop a new method that has the capability to simulate the future electrified and automated freight transportation systems and quantify the national impact of electrified mobility-grid interactions.
    \item Analysis of the impact of truck electrification, automation, and shareability on the grid, infrastructural, fleet, and environmental benefits.
\end{itemize}

The rest of the paper is organized as follows: Section 2 introduces the approaches used for this benefit analysis, Section 3 introduces the extended GEM modeling, Section 4 presents the preliminary results and discusses the results, and finally, Section 5 provides a conclusion.

\section{Approaches}
\label{sec2}

The work developed an optimization model that solves the cost-minimizing dispatch of privately owned and shared heavy-duty vehicles (HDVs) for operation and charging. This optimization model has the capability to study: 1) the allocation of shared heavy-duty autonomous and electric vehicles (SHAEVs) to serve goods -delivery; 2) the investment and construction of a SHAEV fleet and supporting charging infrastructure; and 3) the economic dispatch of electric power plants for the US bulk electricity grid. The power sector was included by coupling GEM to the Grid Operation Optimized Dispatch (GOOD) electricity model (Jenn et al. 2020). This combined model treats the size of the SHAEV fleet and the amount of charging infrastructure as continuous decision variables (relaxing the problem from mixed-integer convex optimization to quadratic programming), allowing for heterogeneous vehicle ranges and charger levels. The model minimizes the total system costs (i.e., operating costs and capital costs) by choosing the timing of vehicle charging subject to several constraints: mobility demand is always served, energy is always conserved, and those generation assets on the grid are dispatched in merit order. Shared autonomous and electric vehicle (SAEV) fleet planning costs are simultaneously minimized by amortizing the cost of the fleet and charging infrastructure to a daily time period. We note that a similar algorithm developed for SAEVs in earlier GEM model developments is incorporated into the formulation for both SHAEVs. 
\begin{figure}[h]
    \centering
    \includegraphics[width = 0.8\linewidth]{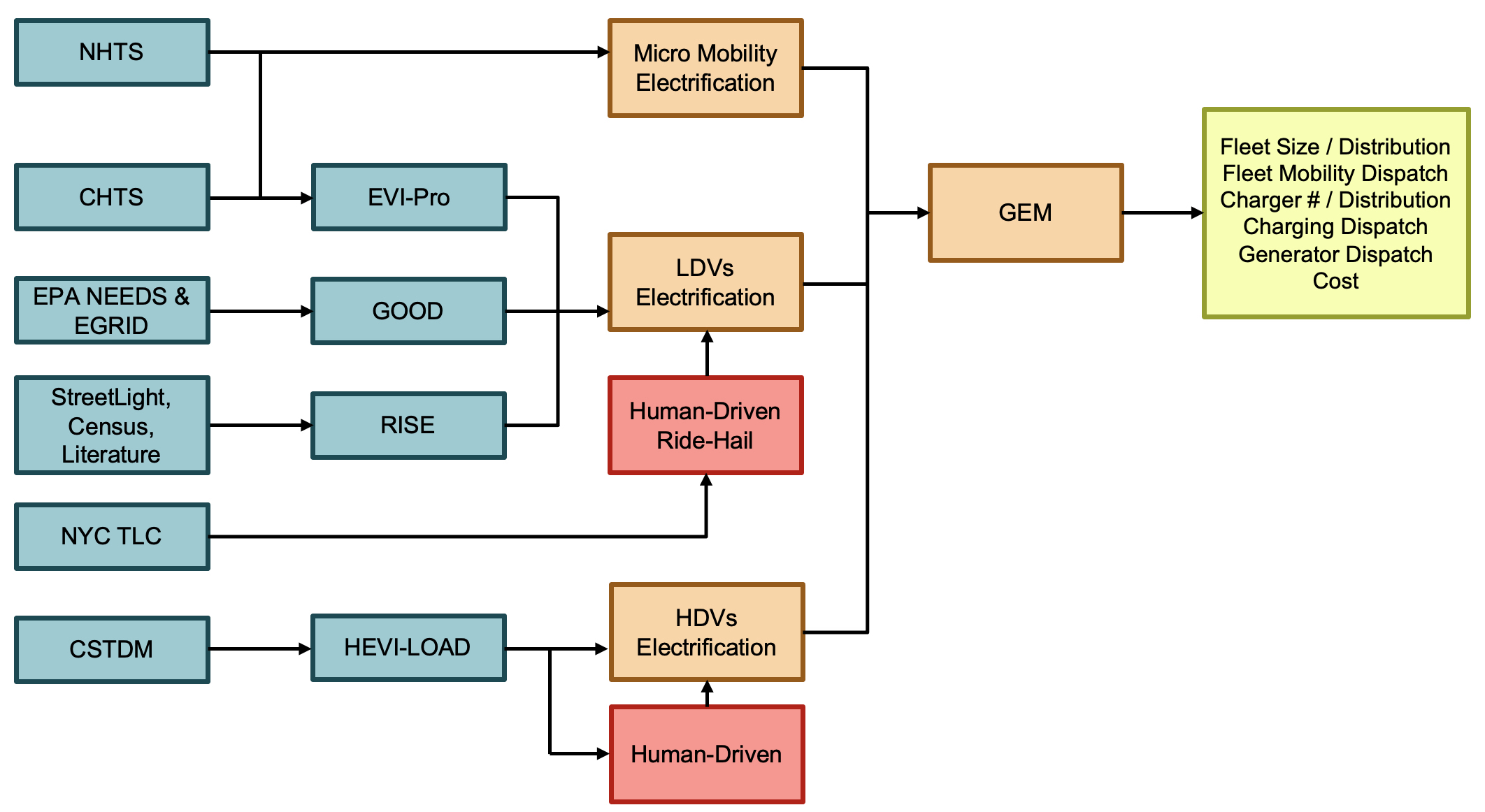}
    \caption{Extended Grid-Integrated Electric Mobility (GEM) model processing workflow.}
    \label{fig:fig1}
\end{figure}

\subsection {GEM}
The Grid-integrated Electric Mobility (GEM) model is an open-source modeling platform developed by researchers at Lawrence Berkeley National Laboratory, UC Davis, and UC Berkeley \cite{sheppard2021grid}. This modeling system simulates the mobility and electricity operations on a national scale. The framework of GEM is unique in that it optimizes a fully autonomous, electric, and shared mobility system while dynamically accounting for high fidelity grid models. The current open-source GEM model is in its first version which mainly focuses on the operation of light-duty vehicles. In this work, we are extending the current GEM model to a broader application that takes the freight behaviors into consideration.  For heavy-duty vehicles, the simulation components are developed using the Medium and Heavy-Duty Electric Vehicle Infrastructure - Load Operations and Deployment (HEVI-LOAD). In this extension, we also added human-driven EV behaviors into the model. Fig. \ref{fig:fig1} shows the extended GEM model framework.

\subsection{HEVI-LOAD}
\begin{figure}[h]
    \centering
    \includegraphics[width = 1\linewidth,height = 0.5\linewidth]{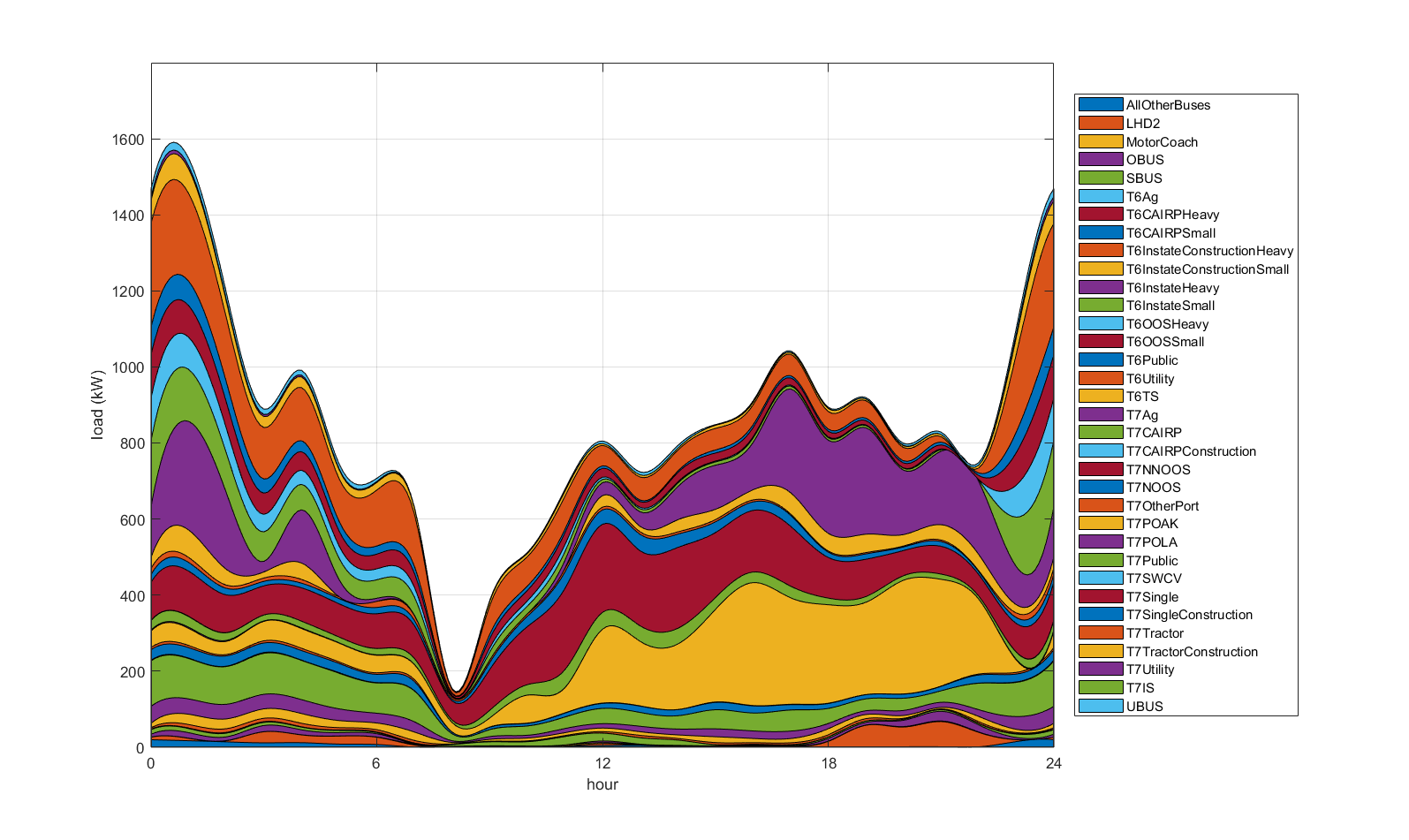}
    \caption{Example Load profile from HEVI-LOAD for California, 2030}
    \label{fig:fig_hevi}
\end{figure}

HEVI-LOAD is a modeling tool developed by Lawrence Berkeley National Laboratory to project the state-wide charging infrastructure needed to accommodate the growing number of medium- and heavy-duty electric vehicles.  To accelerate the decarbonization of medium and heavy-duty (MD/HD) vehicles in California and other states in the United States, HEVI-LOAD projects the number, type, and location of chargers and the related electric grid supply requirements to support the new charging stations. HEVI-LOAD consists of two analytical approaches to determine the load profiles and charging infrastructure needs: 1) the top-down approach that assesses the county-level charging load profile and infrastructure scenarios, and 2) the bottom-up approach that incorporates more granular (temporal, spatial, and duty-cycle-specific) behaviors of a variety of MDHD vehicles into the agent-based activity simulations for optimal charging infrastructure siting and sizing. Figure \ref{fig:fig_hevi} shows the preliminary charging load profile analysis for a variety of MDHDs in California, 2030. 

\section{Problem Formulation}
In the previous GEM model, the light-duty vehicles (LDVs) have been modeled and the optimization problem has been defined under LDV framework \cite{sheppard2021private}. In this work, we have extended the LDV GEM modeling to a more comprehensive optimization model that includes LDVs, HDVs. The dimensions of the model include time, $t$, mobility region $r$, grid region $i$, LDV battery size $b$, HDV battery size $b^H$, LDV charger level $l$, HDV charger level $l^H$, LDV trip distance $d$, HDV trip distance $d^H$, and electricity generator $g$. The model is a quadratically constrained program and can be efficiently solved with a second-order cone programming solver (Cplex).

\subsection{Objective Function} 
As described in the main body of this article and in previous work\cite{sheppard2021private}, the objective function minimizes the amortized daily cost of the fleet and infrastructure, fleet operation, and electricity grid operation.

\begin{equation} \label{obj}
  \mathrm{min}~Z=\sum_r{\left[\sum_t{\left(C^{d}_{tr} + C^m_{tr}\right)}+nC^c_r+nC^v_r\right]} + \sum_{g,t}\left(G_gC^g_g\right) + \sum_{i,t,i'}\left(T_{i,t,i'}C^t_{i,t,i'}\right)
\end{equation}

Where $C^d_{tr}$ is the demand charge or capacity cost to use the grid and $C^m_{tr}$ is vehicle maintenance cost in hour $t$ and mobility region $r$, $C^c_r$ is the amortized daily charging infrastructure cost, $C^v_r$ is the amortized daily fleet cost, $n$ is the number of days in the simulation time horizon, $G_g$ is the electricity produced by generator $g$, $C^g_g$ is the cost of producing a unit of energy by generator $g$, $T_{i,t,i'}$ is the electricity transmitted from grid region $i$ to grid region $i'$, and $C^t_{i,t,i'}$ is the marginal cost of transmission.

The objective is subject to a number of constraints as described in the following section.

\subsection{Constraints}
\textbf{Vehicle Maintenance Cost}: mileage-dependent vehicle maintenance.
\begin{equation} \label{VehicleMaintenanceCost}
  C_{tr}^{m}=\sum_{b,d}{\beta_v V^m_{bdtr} {\nu}_{dtr}}+\sum_{b^H,d^H}{\beta_v^H V^{mH}_{b^Hd^Htr} {\nu}_{d^Htr}^H}
\end{equation}

Where ${\beta}_v,\beta_v^H$ are the per-mile vehicle maintenance costs for LDVs and HDVs, $V^m_{bdtr},V^{mH}_{b^Hd^Htr}$ are the number of vehicles of types $b,b^H$ serving mobility demand of trip length $d,d^H$ in hour $t$ and region $r$, and ${\nu}_{dtr},{\nu}_{d^Htr}^H$ are the average speeds of the vehicles driving trips of length $d,d^H$. Costs associated with cleaning and service are included in maintenance. 
\\
\\
\noindent\textbf{Demand Charge Cost}: cost of grid capacity.

\begin{equation} \label{DemandChargeCost}
  C_{tr}^{d}= P_r^{max} \beta_r / 30.5 / 24
\end{equation}

$P_r^{max}$ is the maximum power demand over the time horizon, ${\beta }_r$ is the average demand charge for the region (\$/kW/month), and 30.5 and 24 convert the monthly demand charge into an hourly value which is summed over all hours in the simulation in the objective function. 
\\
\\
\textbf{Infrastructure Cost}: 

\begin{equation} \label{InfrastructureCost}
  C^c_r=\sum_l{N_{lr}\gamma_l\theta^c_l}+\sum_{l^H}{N^H_{lr}\gamma^H_{l^H}\theta^{cH}_l} 
\end{equation}

Where $N_{lr},N^H_{lr}$ are the number of chargers of power rating $l,l^H$ in region $r$, $\gamma_l$ is the power capacity of the charger (kW), and $\theta^c_l$ is the amortized daily charger cost (\$/kW):

\begin{equation} \label{DailyChargerCost}
  \theta^c_l = \frac{\phi^c_l r (1 + r)^{L^c}}{(1 + r)^{L^c} - 1}
\end{equation}
\begin{equation} \label{DailyChargerCost1}
  \theta^{cH}_l = \frac{\phi^{cH}_l r (1 + r)^{L^{cH}}}{(1 + r)^{L^{cH}} - 1}
\end{equation}

Where $\phi^c_l,\phi^{cH}_l$ are the capital costs of the charger of levels $l,l^H$, $L^c$ is the lifetime of the charger in days, and $r$ is the daily discount rate.
\\
\\
\textbf{Fleet Cost}: in this constraint, battery costs are considered separately from the rest of the vehicle.

\begin{equation} \label{FleetCost}
  C^v_r=\sum_b V^*_{br} (\theta^v + \theta^b B_b)+\sum_{b^H}V^{*H}_{b^Hr} (\theta^{vH} + \theta^{bH} B_b^H)
\end{equation}

Where $V^*_{br}, V^{*H}_{b^Hr}$ are the fleet size for LDVs and HDVs, $\theta^v, \theta^{vH}$ are the amortized daily vehicle costs (without a battery), $\theta^b,\theta^{bH}$ are the amortized daily battery costs (\$/kWh), $B_b,B_b^H$ are the battery capacity (kWh), respectively.

\begin{equation} \label{DailyVehicleCost}
  \theta^v = \psi^{f}_r\left[\phi_{om}^v + \frac{\phi^v r (1 + r)^{L^v}}{(1 + r)^{L^v} - 1}\right]
\end{equation}
\begin{equation} \label{DailyVehicleCost1}
  \theta^{vH} = \psi^{fH}_r\left[\phi_{om}^{vH} + \frac{\phi^{vH} r (1 + r)^{L^{vH}}}{(1 + r)^{L^{vH}} - 1}\right]
\end{equation}

\begin{equation} \label{DailyBatteryCost}
  {\theta^b = \psi^{b}_r\left[\frac{\phi^b r (1 + r)^{L^b}}{(1 + r)^{L^b} - 1}\right]}
\end{equation}
\begin{equation} \label{DailyBatteryCost1}
  \theta^{bH} = \psi^{bH}_r\left[\frac{\phi^{bH}r (1 + r)^{L^{bH}}}{(1 + r)^{L^{bH}} - 1}\right]
\end{equation}

Where $\psi^{f}_r,\psi^{fH}_r$ are the fleet spatial mismatch correction factors (see Bauer et al.\cite{bauer_cost_2018}), $\phi_{om}^v,\phi_{om}^{vH}$ are the daily fixed O\&M costs for the vehicle, $\phi^v,\phi^{vH}$ are the capital costs of the vehicles, and $L^v,L^{vH}$ are the lifetimes of the vehicles in days. And where $\psi^{b}_r,\psi^{bH}_r$ are the battery spatial mismatch correction factors (see Bauer et al.\cite{bauer_cost_2018}), $\phi^b,\phi^{bH}r$ are the capital costs of the battery (\$/kWh) and $L^b,L^{bH}$ are the lifetimes of the battery in days.
\\
\\
\textbf{Demand Allocation}: mobility demand must be served by some composition of vehicles.

\begin{equation} \label{DemandAllocation}
  \sum_b{D_{bdtr}}={DD}_{dtr}
\end{equation}
\begin{equation} \label{DemandAllocation1}
  \sum_{b^H}{D_{b^Hd^Htr}^H}={DD}_{d^Htr}^H
\end{equation}

Where ${DD}_{dtr},{DD}_{d^Htr}^H$ is exogenous demands in hour $t$ for passenger vehicles (LDVs) and trucks (HDVs).
\\
\\
\textbf{Energy to Meet Demand}: the energy consumed by the fleet is a function of the number of trips served, and the conversion efficiency of the vehicles. We model the effect of urban form and sharing as multipliers on the energy efficacy of serving mobility demand.

\begin{equation} \label{EnergyToMeetDemand}
E_{bdtr} =\frac{D_{bdtr} \psi^{chdd}_r \psi^{cdd}_r \eta_b\rho_d}{\sigma_d}
\end{equation}

\begin{equation} \label{EnergyToMeetDemand1}
E^H_{b^Hd^Htr} =\frac{D^H_{b^Hd^Htr} \psi^{chddH}_r \eta^H_{b^H}\rho^H_{d^H}}{\sigma^H_{d^H}}
\end{equation}

Where $E_{bdtr},E^H_{b^Hd^Htr}$ are the energy consumed serving mobility of vehicle types $b,b^H$ and trip length $d,d^H$ in hour $t$ and region $r$, ${\sigma }_d,\sigma^H_{d^H}$ are the sharing factor or the average number of passengers per vehicle trip, and the sharing factor for shared automated trucks mapped to per vehicle trip from private trucks, $\psi^{chdd}_r,\psi^{chddH}_r$ are the charge deadhead distance correction ratios (see \cite{bauer_cost_2018}),  $\psi^{cdd}_r$ is the customer deadhead distance correction ratios, and ${\eta }_b,\eta^H_{b^H}$ is the conversion efficiency of the vehicle power trains of LDVs and HDVs (kWh/mile).
\\
\\
\textbf{Vehicles Moving}: the number of vehicles actively serving trips is related to trip demand and the sharing factor. The terms $\frac{{\rho }_d}{\Delta t{\nu }_{dt}}$, $\frac{\rho_{d^H}^H}{\Delta t\nu_{d^Htr}^H}$ correct for the length of the time period, allowing, e.g. 1 vehicle to serve 2 trips in an hour if the distance to speed ratio is 1/2.

\begin{equation} \label{NumMoving}
V^m_{bdtr}=\frac{D_{bdtr}\rho_d\psi^{cdt}_r}{\sigma_d\Delta t\nu_{dtr}}
\end{equation}

\begin{equation} \label{NumMoving1}
V^{mH}_{b^Hd^Htr}=\frac{D_{b^Hd^Htr}^H\rho_{d^H}^H}{\sigma_{d^H}^H\Delta t\nu_{d^Htr}^H}
\end{equation}

Where $\psi^{cdt}_r$ is the customer deadhead time correction ratio, and $\Delta t$ is the length of the time period in hours. 
\\
\\
\textbf{Vehicles Charging}: we relate the number of vehicles charging to the power consumed by the capacity of each charger type.

\begin{equation} \label{NumCharging}
V^c_{btlr}=\frac{P_{btlr}}{\psi^{chdt}_{b,l,r}{\gamma }_l}
\end{equation}
\begin{equation} \label{NumCharging1}
V^{cH}_{b^Htl^Hr}=\frac{P_{b^Htl^Hr}^H}{\psi^{chdtH}_{b^H,l^H,r}{\gamma }_{l^H}^H}
\end{equation}

Where $V^c_t$ are the number of vehicles charging in hour $t$, $\psi^{chdt}_{b,l,r},\psi^{chdtH}_{b^H,l^H,r}$ are the charger deadhead time correction ratios, and ${\gamma }_l,\gamma_{l^H}^H$ are the charging rates (kW / charger).
\\
\\
\textbf{Charging Upper Bound}: we assume the batteries in the fleet start full and therefore can only be replenished up to the cumulative amount consumed by the previous hour. 

\begin{equation} \label{ChargingUpperBound}
\sum^t_{\hat{t}=0}{\sum_l{P_{b\hat{t}lr}}}\le \sum^{t-1}_{\hat{t}=0}{\sum_d{E_{bd\hat{t}r}}}, ~~~~ \forall  btr
\end{equation}
\begin{equation} \label{ChargingUpperBound1}
\sum^t_{\hat{t}=0}{\sum_{l^H}{P^H_{b^H\hat{t}l^Hr}}}\le \sum^{t-1}_{\hat{t}=0}{\sum_d{E^H_{b^Hd^Htr}}}, ~~~~ \forall  b^Htr
\end{equation}

\textbf{Charging Lower Bound}: charging must keep up with consumption as limited by the capacity of the batteries. Energy must be supplied by charging in the previous hour to be used in the next hour. 

\begin{equation} \label{charginglowerbound}
  \sum^{t-1}_{\hat{t}=0}{\sum_l{P_{b\hat{t}lr}}} \ge \sum^t_{\hat{t}=0}{\sum_d{E_{bd\hat{t}r}}}-V^*_{br}B_b, ~~~~ \forall  btr
\end{equation}
\begin{equation} \label{charginglowerbound1}
  \sum^{t-1}_{\hat{t}=0}{\sum_{l^H}{P^H_{b^H\hat{t}l^Hr}}} \ge \sum^t_{\hat{t}=0}{\sum_{d^H}{E^H_{bd\hat{t}r}}}-V^{*H}_{b^Hr}B_{b^H}^H, ~~~~ \forall  b^Htr
\end{equation}

\textbf{No Charge At Start}: the first hour of the day needs to have no charging to allow for the convention that charging can only occur after some energy is consumed by the fleet. 

\begin{equation} \label{NoChargeAtStart}
P_{btlr}=0, t=0, ~~~~ \forall  blr
\end{equation}
\begin{equation} \label{NoChargeAtStart1}
P^H_{b^Htl^Hr}=0, t=0, ~~~~ \forall  b^Hl^Hr
\end{equation}
\\
\textbf{Terminal State of Charge}: the aggregate state of charge of batteries must again be full at the end of the simulation. 

\begin{equation} \label{TerminalSOC}
\sum_t{\sum_l{P_{btlr}}}=\sum_t{\sum_d{E_{bdtr}}}, ~~~~ \forall  br
\end{equation}
\begin{equation} \label{TerminalSOC1}
\sum_t{\sum_{l^H}{P^H_{b^Htl^Hr}}}=\sum_t{\sum_{d^H}{E^H_{b^Hd^Htr}}}, ~~~~ \forall  b^Hr
\end{equation}
\\
\noindent\textbf{Fleet Dispatch}: together vehicles serving trips, charging, and idle cannot exceed the fleet size.

\begin{equation} \label{FleetDispatch}
  \sum_d{V^m_{bdtr}}+V^i_{btr}+\sum_l{V^c_{btlr}}\le V^*_{br}
\end{equation}
\begin{equation} \label{FleetDispatch1}
  \sum_{d^H}{V^{mH}_{b^Hd^Htr}}+V^{iH}_{b^Htr}+\sum_l{V^{cH}_{b^Htl^Hr}}\le V^{*H}_{b^Hr}
\end{equation}
\\
\textbf{Max Charging}: vehicles charging cannot exceed the number of chargers.

\begin{equation} \label{MaxCharging}
  \sum_{bd}{V^c_{bdtl}}\le N_{lr}
\end{equation}
\begin{equation} \label{MaxCharging1}
  \sum_{b^Hd^H}{V^{cH}_{b^Hd^Htl^H}}\le N_{l^Hr}^H
\end{equation}

Where $N_{lr}$ is the number of chargers charging at power level $l$ in region $r$.
\\
\\
\textbf{Max Demand}: this constraint relates the maximum power consumed for each region to the power drawn in each time period. Because $P_r^{max}$ is in the objective function, there will be no slack in the optimal solution, ensuring it will be equal to the maximum power demanded by the fleet.

\begin{equation} \label{MaxDemand}
  P_r^{max} \ge \frac{\sum_{bl}P_{tblr}}{\Delta t} +\frac{\sum_{b^Hl^H}P_{tb^Hl^Hr}^H}{\Delta t}- P^{private}_{t,r} - P^{H private}_{t,r},-P^{H hdr}_{t,r} ~~~~ \forall tr
\end{equation}

Where $P^{private}_{t,r}$ is the power demanded by the personally owned light-duty EV fleet, $P^{H private}_{t,r},P^{H hdr}_{t,r}$ are the power demanded by the personally owned heavy-duty EV fleet
\\

\noindent\textbf{Personal HDV Charging (automated)}: The light-duty vehicle personal vehicle charging constraints are derived in our previous work \cite{sheppard2021private}. The following four constraints represent the power and energy bounds on personal HDV charging.

\begin{eqnarray} \label{PersonalEVChargeConstraints}
  P^{H private}_{t,r} \ge \underbar{P}^{H private}_{t,r} \\
  P^{H private}_{t,r} \le \overline{P}^{H private}_{t,r} \\
  \sum^t_{t'=1}{P^{H private}_{t',r}} \ge \underbar{E}^{H private}_{t,r} \\
  \sum^t_{t'=1}{P^{H private}_{t',r}} \le \overline{E}^{H private}_{t,r}
\end{eqnarray}

Where $\underbar{P}^{H private}_{t,r}$ and $\overline{P}^{H private}_{t,r}$ are the min and max power constraints on EV charging, respectively; and $\underbar{E}^{H private}_{t,r}$ and $\overline{E}^{H private}_{t,r}$ are the min and max cumulative energy constraints on EV charging, respectively.
\\

\noindent\textbf{Personal HDV Charging (human-driven)}: the following four constraints represent the power and energy bounds for the HDV human-driven charging behavior.

\begin{eqnarray} \label{PersonalEVChargeConstraints}
  P^{H hdr}_{t,r} \ge \underbar{P}^{H hdr}_{t,r} \\
  P^{H hdr}_{t,r} \le \overline{P}^{H hdr}_{t,r} \\
  \sum^t_{t'=1}{P^{H hdr}_{t',r}} \ge \underbar{E}^{H hdr}_{t,r} \\
  \sum^t_{t'=1}{P^{H hdr}_{t',r}} \le \overline{E}^{H hdr}_{t,r}
\end{eqnarray}

Where $\underbar{P}^{H hdr}_{t,r}$ and $\overline{P}^{H hdr}_{t,r}$ are the min and max power constraints on the human driven HDV charging, respectively; and $\underbar{E}^{H hdr}_{t,r}$ and $\overline{E}^{H hdr}_{t,r}$ are the min and max cumulative energy constraints on the human driven HDV charging, respectively.
\\
\\
\noindent\textbf{Generation}: The following three constraints represent power generation on the grid.

\begin{eqnarray} \label{Generation}
  \sum_{g,i}G_{g,t} + \eta^{trans}\sum_{i'}T_{i',t,i} - \sum_{i'}T_{i,t,i'} \ge &P^{other}_{i,t} + \sum_{r\epsilon i}P^{private}_{t,r} + \sum_{r\epsilon i,b,l}P_{tblr}+ \sum_{r\epsilon i}P^{H private}_{t,r}\nonumber \\& +\sum_{r\epsilon i}P^{H hdr}_{t,r} + \sum_{r\epsilon i,b^H,l^H}P^H_{tb^Hl^Hr}
\end{eqnarray}

For all time steps $t$ and grid regions $i$, where $P^{other}_{t,r}$ is electricity demand from non-mobility sources, and $\eta^{trans}$ is the transmission loss factor associated with inter-regional transfers.

\section{Preliminary results}
In this section, the simulation results are presented and the benefit analysis is given based on the simulation study via the GEM model.

\begin{figure}[h]
    \centering
    \subfigure[HDV charging load profile in nation-wide mobility regions]{\includegraphics[width=0.4\textwidth,height = 0.3\textwidth,trim={0 0 180 0},clip]{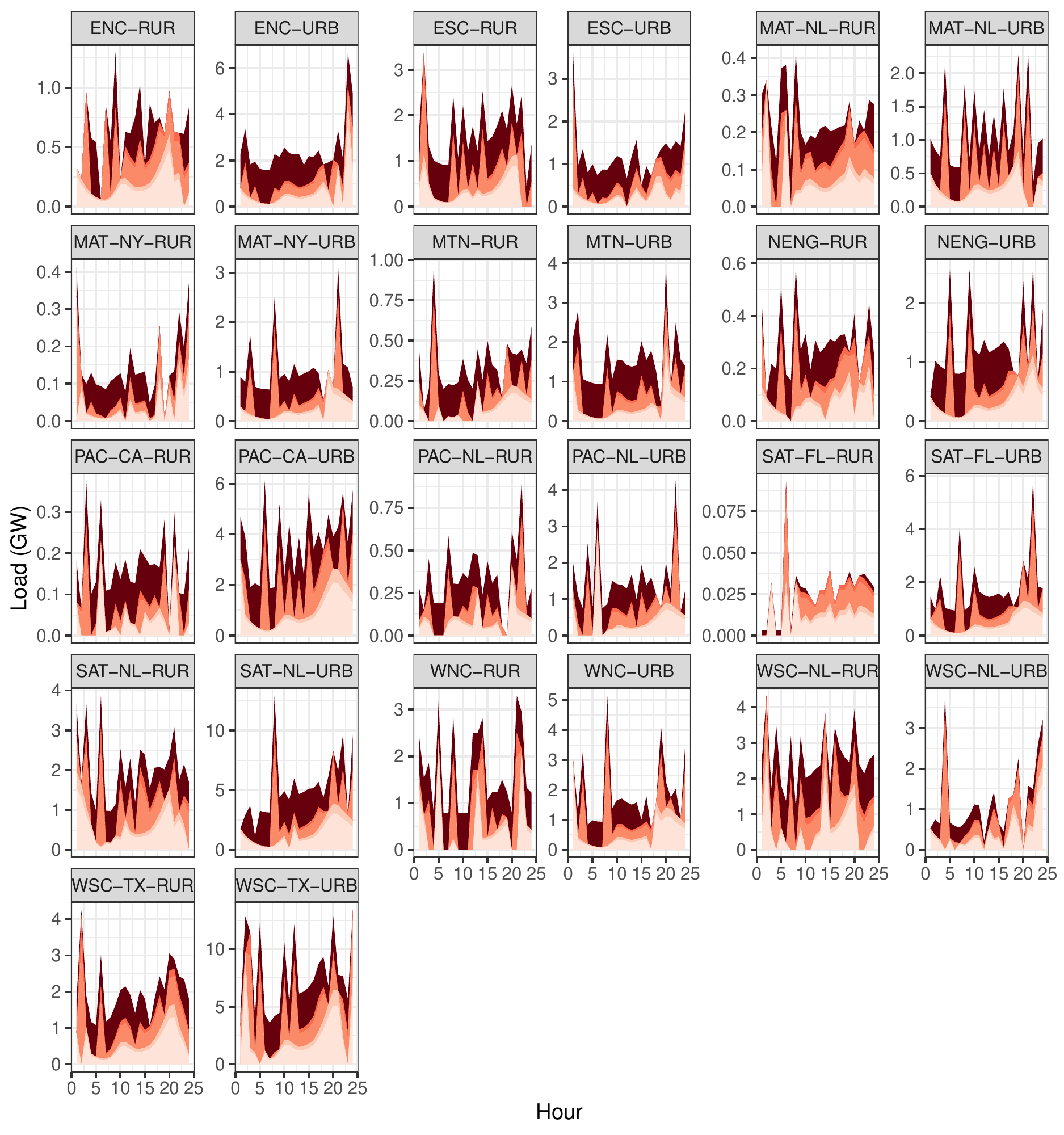}\label{fig:sub11}} 
    \subfigure[Nation-wide HDV charging load profile]{\includegraphics[width=0.55\textwidth]{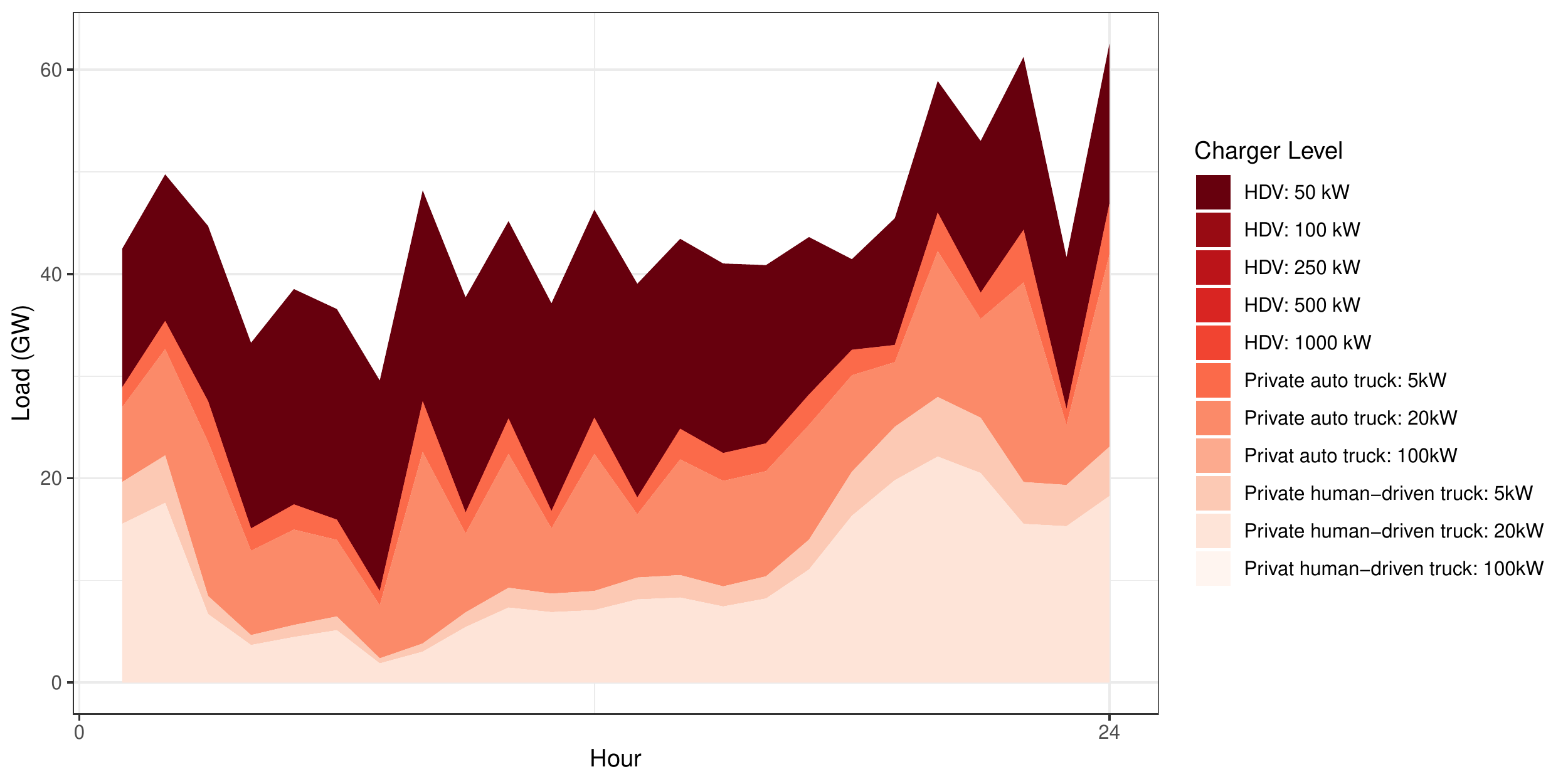}\label{fig:sub12}} 
    \caption{Electrified HDV charging load profile}
    \label{fig:load_profile}
\end{figure}

\textbf{HDV charging load profile}. Figure \ref{fig:load_profile} shows the overall charging load profile for all kinds of electrified trucks including SHAEVs, private owned automated trucks, and private-owned human-driven trucks with the use of different charging levels. We assume for all the electrified trucks, 50\% of them are SHAEVs ($S = 50\%$), and 50\% of them are private owned fleets ($P = 50\%$). Among the privately owned fleets, 50\% of the fleets are privately owned automated trucks, and the rest 50\% of the fleets are privately owned human-driven trucks.

\begin{figure}[h]
    \centering
    \subfigure[Numbers of chargers needed]{\includegraphics[width=0.48\textwidth]{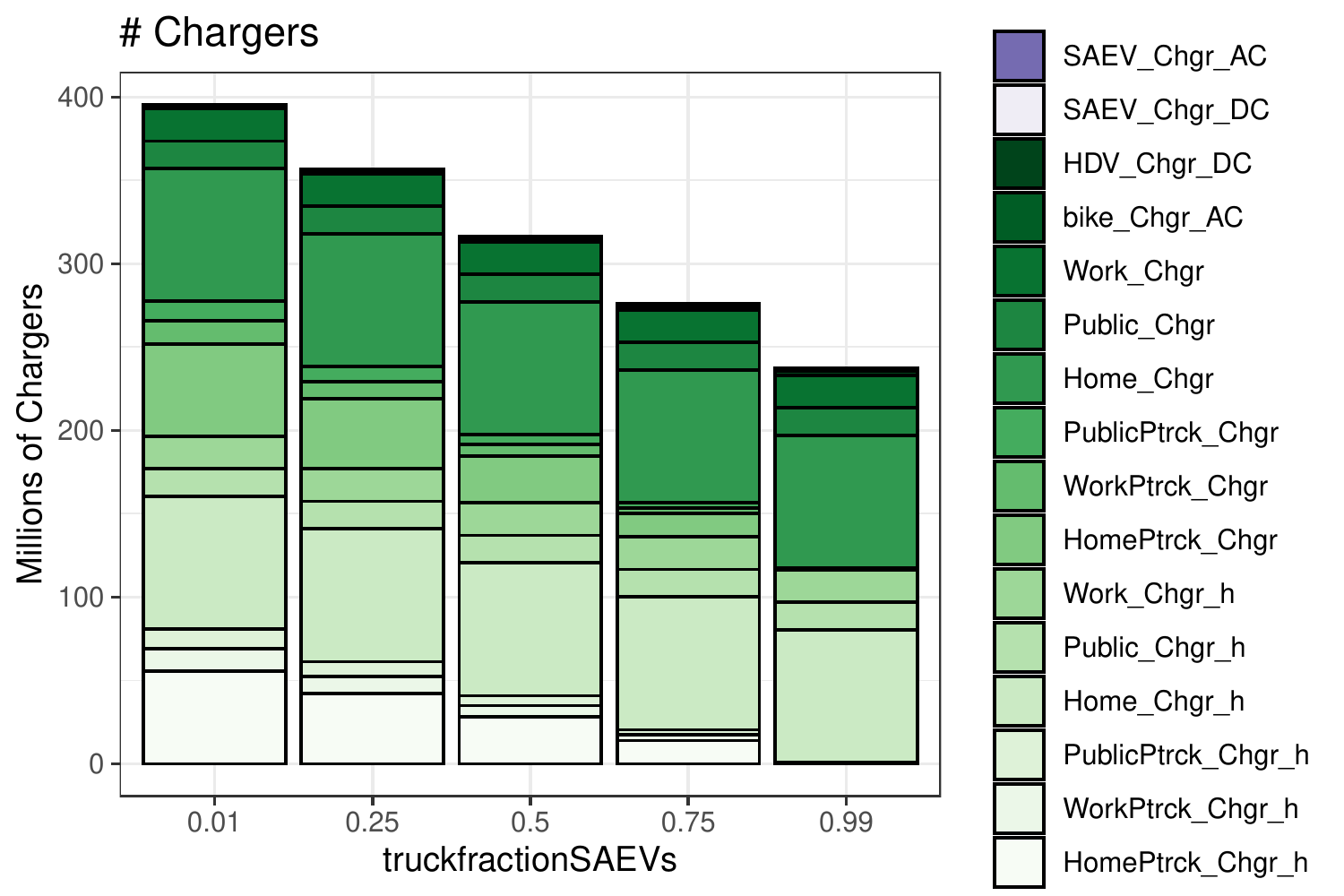}\label{fig:sub21}} 
    \subfigure[Peak load]{\includegraphics[width=0.48\textwidth]{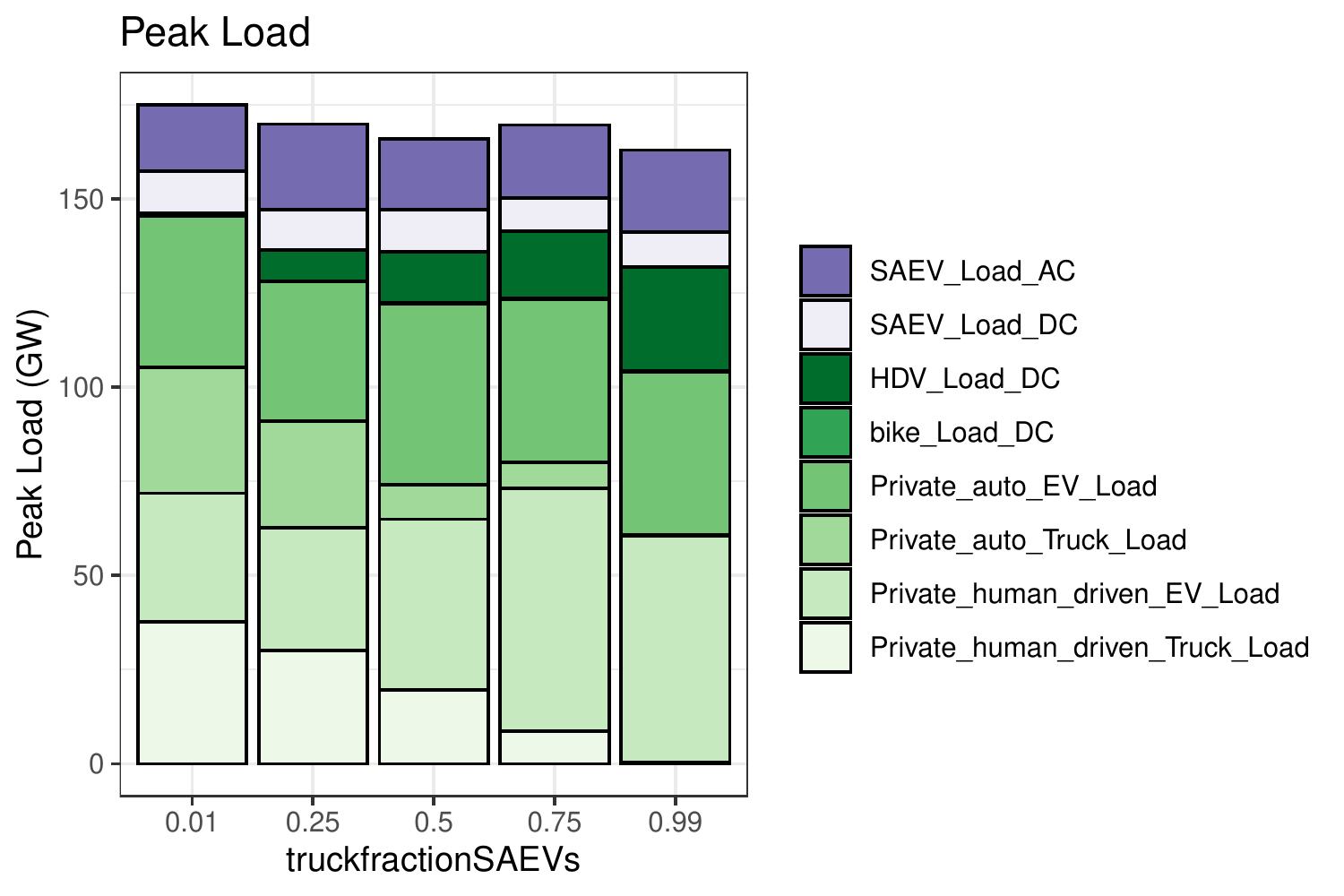}\label{fig:sub22}} 
    
    \subfigure[Fleet size]{\includegraphics[width=0.48\textwidth]{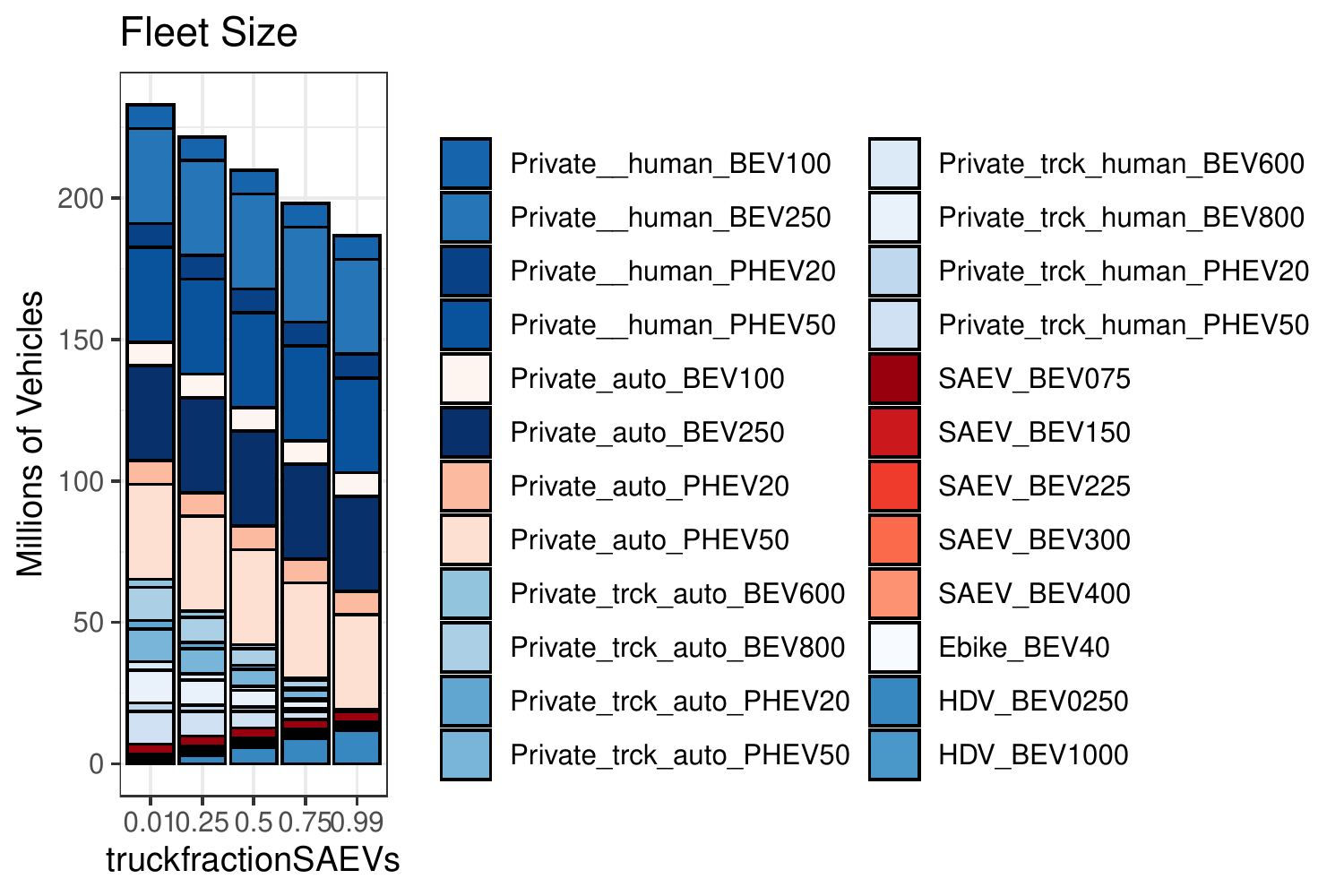}\label{fig:sub23}} 
    \subfigure[Overall cost]{\includegraphics[width=0.48\textwidth]{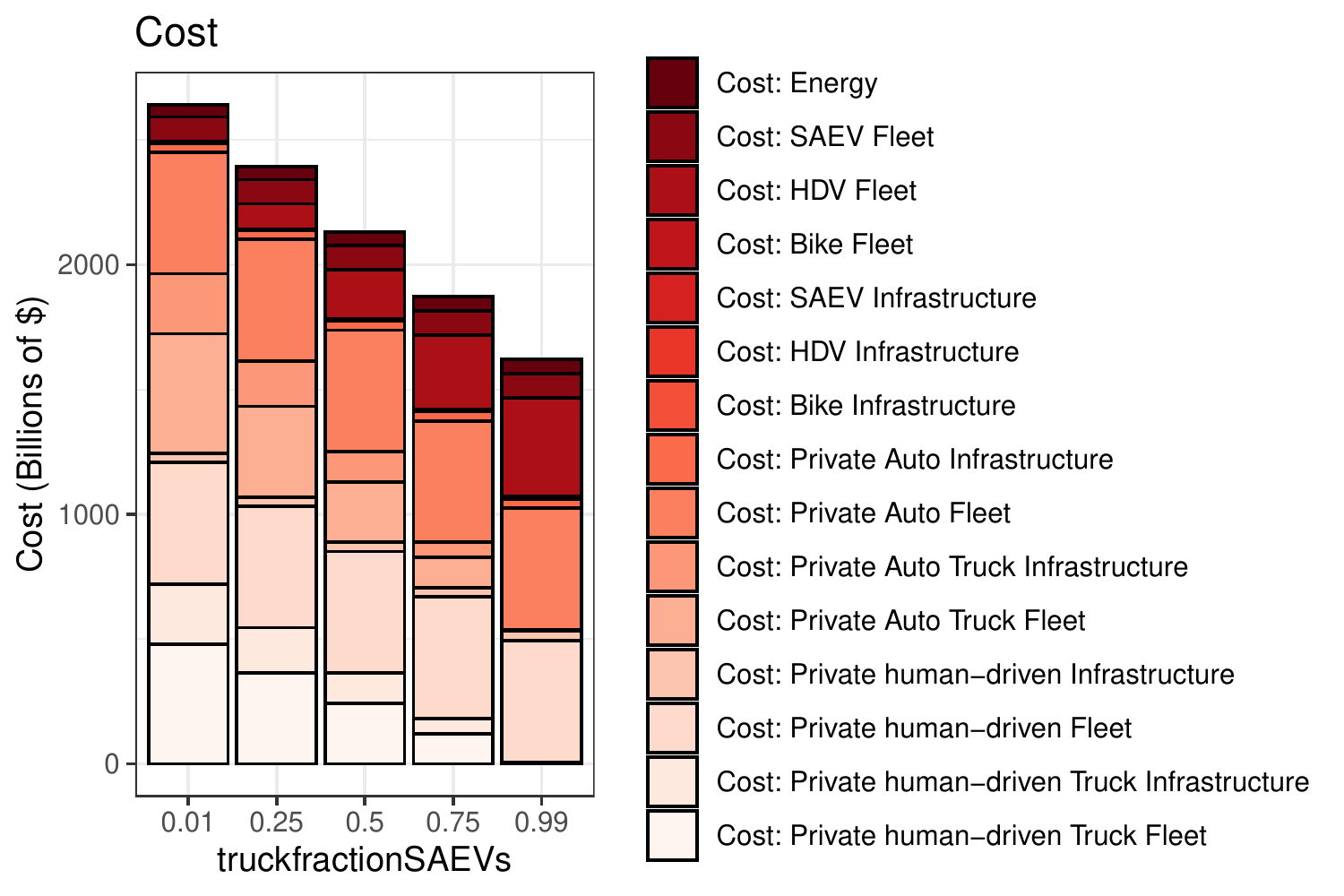}\label{fig:sub24}} 
    \caption{Numbers of chargers}
    \label{fig:analysis}
\end{figure}

\textbf{Number of chargers}. Figure \ref{fig:sub21} shows the number of chargers needed. As with fleet size, there are far more chargers when SHAEVs are not available (S = 0\%) than in a counterfactual scenario of a fully SHAEV fleet (S = 100\%), reflecting much higher utilization among SHAEV chargers.

\textbf{Peak load}. Figure \ref{fig:sub22} shows the grid peak load, which also decreases substantially as the fraction of mobility demand met by SHAEVs increases: Peak demand is 176 GW at S = 0\% and is 165 GW when S = 100\%. 

\textbf{Fleet size}. Figure \ref{fig:sub23} shows the optimal fleet size of all types of vehicles in GEM modeling. We are particularly investigating the SHAEVs and privately-owned electrified HDVs in this study which decreases 105M vehicles from the S = 0\% case to the S = 100\% case. 

\textbf{Total costs}. Figure \ref{fig:sub24} shows the overall cost changes with the fraction of SHAEVs increases. We can observe that the fleet cost and infrastructure cost for the privately owned electrified HDVs are decreasing with a larger scale compared to the increment of fleet cost and infrastructure cost related to the increase of SHAEV fleets.

\subsection{Discussion}

With the growing trend of freight electrification, there is an urgent need to understand the economic and environmental benefits of future freight components. With the simulation results, we find that: 1)The use of shared heavy-duty autonomous electric vehicles (SHAEVs) to provide goods delivery has substantial benefits over using privately-owned trucks or gasoline trucks; 2)Without charging time requirements, lower power charging stations and use of smaller battery size trucks provide the maximum benefit in terms of infrastructure and fleet cost, and grid benefit.

\section{Conclusion}

The configuration of the freight system in which SHAEVs serve goods delivery has substantial benefits over one that relies on privately-owned electrified trucks or gasoline-powered vehicles. Overall, we demonstrate that freight automation increases operating efficiency by reducing total costs and lowering emissions, which also increases goods delivery within the transportation system. From an economic standpoint, system costs are substantially reduced through sharing and automation, while fuel and operational costs remain much lower than those of gasoline/diesel vehicles today. From an electric power grid operator’s perspective, SHAEVs can smooth out large amounts of the variability in electricity generation, which substantially improves both the efficiency and emissions rate of fossil generation while simultaneously better utilizing solar and wind resources (thanks to the flexibility in charging times).

%Bibliography
\bibliographystyle{unsrt}  
\bibliography{references}

\end{document}